# Synthesis of Long-T$_1$ Silicon Nanoparticles for Hyperpolarized $^{29}$Si Magnetic Resonance Imaging

*Tonya M. Atkins,[1] Maja C. Cassidy,[2] Menyoung Lee,[3] Shreyashi Ganguly,[1] Charles M. Marcus[3,4]\* and Susan M. Kauzlarich[1]\**

[1]Department of Chemistry, University of California, Davis, CA 95616 USA
[2]School of Engineering and Applied Science, Harvard University, Cambridge, MA 02138 USA
[3]Department of Physics, Harvard University, Cambridge, MA 02138 USA
[4]Center for Quantum Devices, Niels Bohr Institute, University of Copenhagen, 2100 Copenhagen O, Denmark

ABSTRACT

We describe the synthesis, materials characterization and dynamic nuclear polarization (DNP) of amorphous and crystalline silicon nanoparticles for use as hyperpolarized magnetic resonance imaging (MRI) agents. The particles were synthesized by means of a metathesis reaction between sodium silicide (Na$_4$Si$_4$) and silicon tetrachloride (SiCl$_4$) and were surface functionalized with a variety of passivating ligands. The synthesis scheme results in particles of diameter ~10 nm with long size-adjusted $^{29}$Si spin lattice relaxation (T$_1$) times (> 600 s), which are retained after hyperpolarization by low temperature DNP.

Silicon nanoparticles (Si NPs) with sizes ranging from 1 nm to 100 nm are of great interest across many research areas including thermoelectrics, photovoltaics, nanoelectronics and nanomedicine. The smaller sizes (≤ 5 nm) emit light due to quantum confinement effects[1-3] while larger sizes (≥ 10 nm) have desirable electronic transport properties and are pursued for the production of photovoltaics.[4, 5] A variety of methods have been investigated for the production of Si NPs. Use of thermal techniques such as laser pyrolysis and thermal decomposition has been widespread,[1, 6-10] but control over size and material properties using these methods is poor. Synthetic colloidal chemistry allows the simultaneous control of both the size and surface of the nanoparticle, offering prospects of creating large quantities of high purity monodisperse Si NP samples. Previously, small (1-2 nm diameter) Si NPs have been prepared by the reaction of metal silicides with silicon tetrachloride.[11, 12] The reaction of the Zintl salt with ammonium halide has also been reported with average Si NP sizes of 4 nm.[13, 14] Si NPs with a wide size distribution (1-20 nm) have also been prepared by a reverse micelle process both at ambient conditions and under high temperatures and pressures in a bomb.[15, 16] Solution synthesis of Si NPs with diameters greater than 10 nm with discrete size distributions is challenging, and there are few examples reported in literature.[17, 18] Recently, size control in the quantum confinement regime (≤ 6 nm) has been demonstrated through a combination of high temperature treatment and etching with HF.[19] Alternative solution routes need to be explored to attain discrete size ranges for use in several applications.

Si and its oxide derivatives have emerged as materials of choice for applications in nanomedicine due to their biocompatibility and biodegradability *in-vivo*,[20, 21] and are under investigation as drug delivery vehicles[22-26] and imaging agents.[27-33] Oxide derivatives, such as Stöber and fumed silicas, are not created equally and toxicity has been shown to differ depending on the silica structure and is also dependent on the surface chemistry, reactivity and NP morphology.[34] *In-vivo* imaging of Si NPs has been achieved *via* a variety of techniques including optical imaging,[21, 22, 27, 28] positron emission tomography (PET)[29] and magnetic resonance imaging (MRI).[35] MRI is an attractive technique for non-invasive imaging because it allows high-contrast imaging of opaque structures with detailed anatomical resolution without the use of ionizing radiation. *In-vivo* MRI of Si particles has previously been performed through the incorporation of paramagnetic materials[24, 27, 28] that disrupt the relaxation and coherence times of proximal $^1$H nuclei, resulting in negative contrast, or a lack of signal on a $^1$H anatomical image. Recently, a new technique has been demonstrated for positive-contrast imaging of silicon particles using MRI[32, 33] where the $^{29}$Si nuclei within silicon particles (4.7 % natural abundance) are hyperpolarized and imaged directly *in-vivo* using $^{29}$Si MRI. The body naturally contains very little silicon so this technique allows for positive contrast background-free imaging, similar to PET, but without the use of ionizing radiation. The initial demonstrations of $^{29}$Si hyperpolarized MRI utilized commercially sourced micron-scale particles that could be monitored in the *in-vivo* environment for over 40 minutes.[32] In intravenous studies, these particles were confined to the vascular structure of the animal, as their size rendered them unable to cross the cell wall.

At standard experimental temperatures and magnetic fields the nuclear spin polarization is extremely low, on the order of one part in $10^{-5}$ – $10^{-7}$ depending on the nuclear species. Hyperpolarization increases the nuclear spin polarization by several orders of magnitude, allowing the investigation of weak nuclear species by nuclear magnetic resonance (NMR) in low concentrations.[36] Hyperpolarization has been used in materials science as a nanoscale characterization tool,[37, 38] while in biochemistry it has been used to reveal the structure of complex biomolecules such as proteins and peptides.[39, 40] In clinical applications, hyperpolarized noble gases have allowed detailed imaging of the lung[41] and metabolites such as $^{13}$C labeled pyruvate have been used for monitoring therapy in oncology.[42, 43]

The spin-lattice relaxation time ($T_1$) sets the approximate time scale for which the signal from the hyperpolarized nuclei can be observed for imaging. Most enriched metabolites and noble gases have $T_1$ times on the order of 5 - 30 s in the *in-vivo* environment, as the nuclei are strongly coupled to local $^1$H spins.[41-43] Although a range of impressive *in-vivo* experiments have been undertaken with these materials,[41-43] these times are at the very lowest of what is feasible for imaging clinically relevant biological processes. Extending the $T_1$ times of hyperpolarized tracers to > 10 minutes would make a wide range of biological processes accessible for imaging using the high specificity and resolution of MRI.

In solid-state spin-½ materials, such as Si, the primary source of nuclear spin-lattice relaxation is *via* an interaction with unbonded electrons or holes which occur in the form of free carriers or paramagnetic centers existing at lattice defect sites or at the surface. If the material is pure, the nuclear spin polarization is largely protected from sources of relaxation, resulting in bulk relaxation times of many hours.[32, 36, 44] Previous NMR investigations of Si micro- and nanoparticles showed that reducing the bulk to the nanoscale results in $T_1$ times that are strongly dependent on the particle size, purity and density of isolated defects.[32] The $^{13}$C nucleus in diamond is another solid-state spin ½ system under investigation as a hyperpolarized imaging agent, however to date measured $T_1$ times have been extremely short, less than 2 s for particles 5 - 30 nm in diameter.[45]

The colloidal synthesis of Si NPs with average diameters of ~10 nm with room temperature $T_1$ times that exceed 600 s is described. The silicon nanoparticles were prepared from the metathesis reaction between sodium silicide ($Na_4Si_4$) and silicon tetrachloride ($SiCl_4$). While small (≤ 5 nm) NPs can be prepared by this route,[11, 12, 46] this work presents the larger NPs that can also be obtained from this reaction, isolated by means of product precipitation and density differences of the precipitate as outlined in the Methods section.[47, 48] This results in particles that have significantly longer size-adjusted $^{29}Si$ nuclear spin relaxation ($T_1$) times than other commercially sourced Si particles,[32] as well as diamond nanoparticles of similar sizes.[45] We show that all of these particles can be hyperpolarized by low temperature dynamic nuclear polarization (DNP), with polarization dynamics similar to the micron-scale Si particles previously investigated. These Si NPs are the smallest studied to date for hyperpolarization and, due to their size and long $T_1$ times, are promising for use as *in-vivo* imaging agents that are small enough for intracellular studies and to cross the blood-brain barrier.[49]

RESULTS AND DISCUSSION

We used a metathesis reaction of sodium silicide ($Na_4Si_4$) with a silicon tetrachloride ($SiCl_4$) in each reaction for forming the chlorinated Si NPs (Figure 1).[11, 12, 46] The chlorinated Si NPs were then surface passivated with either an alkane or an aromatic amine by reacting them with the Grignard reagents, octylmagnesium bromide ($C_8H_{17}MgBr$) (Figure 1a) or 4-aminophenylmagnesium bromide ($C_6H_6NMgBr$) (Figure 1b), in dimethoxyethane (DME) or dioctyl ether (DOE)/squalane mixture. The ligands, octyl- and aminophenyl-, were chosen to provide both a nonpolar and polar surface group. Si NPs with average sizes of ~10 nm were isolated by collecting the portion of the precipitated product that remained at the interface between the water and organic layers. To enhance the crystallinity of the Si NPs, a third reaction was run (Figure 1c) where the chlorinated Si NPs were annealed in forming gas (95 % Ar, 5% H$_2$) at 650° for 20 mins and subsequently passivated with aminopropyltrimethoxy silane (APTMS).

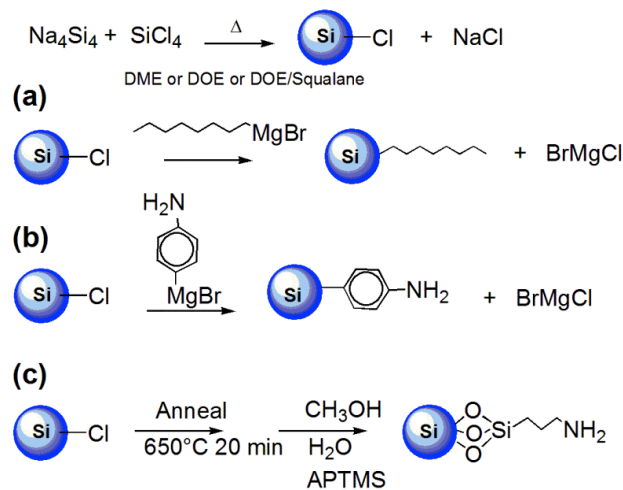

**Figure 1.** Synthetic schemes used to produce Si NPs. Each reaction starts with the reaction of the Zintl salt, Na$_4$Si$_4$ with SiCl$_4$ to form the chlorinated Si NPs, and then reacted with (a) octylmagnesium bromide, (b) 4-aminophenylmagnesium bromide and (c) annealed and passivated with aminopropyltrimethoxysilane (APTMS).

Transmission electron microscope (TEM) images of Si NPs produced according to the schemes in Figure 1(b) and (c) are shown in Figure 2. Size distributions based on a number of images from these reactions are 10.3 ± 3.2 nm for the 4-aminophenyl terminated Si NPs and 10.4 ± 4.4 nm for the annealed APTMS terminated NPs. The images shown are representative of the products obtained according to the reactions outlined in Figure 1. The octyl terminated Si NPs described by Figure 1(a) have a similar size and size distribution to that of the 4-aminophenyl terminated NPs. All reaction schemes produced dispersed Si NP samples that suspend well in solution and were free from aggregation, indicating good termination of the particle surface. The annealed Si NPs showed a visible diamond crystalline lattice structure under high resolution TEM (HRTEM) (inset, Figure 2b) and showed no additional aggregation after annealing due to sintering.

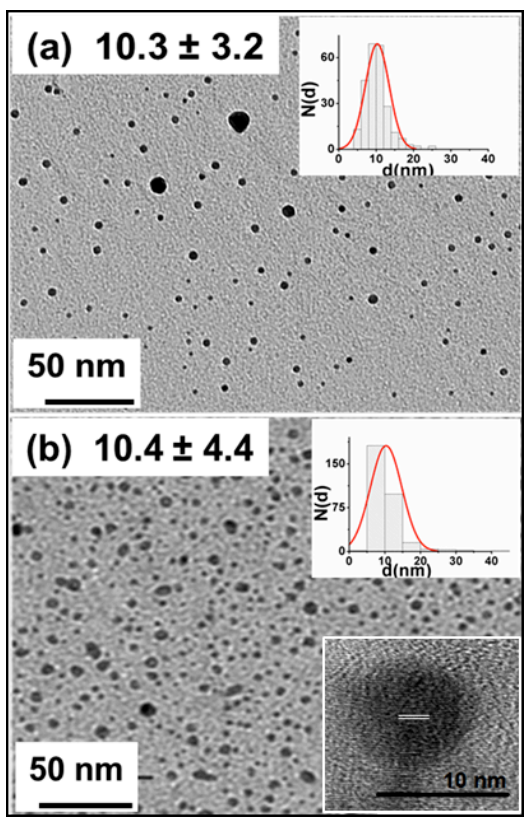

**Figure 2.** TEM images of Si NPs: (a) 4-aminophenyl terminated and (b) annealed and passivated with APTMS. The size distributions are shown in the upper right inset. The lower right inset in Figure 2b shows the crystallinity of the annealed Si NPs obtained using HRTEM. The lines in the lower right inset indicate the lattice fringes which give rise to the (220) lattice plane for diamond structured Si.

X-ray powder diffraction (XRD) and Raman spectroscopy were used to determine the degree of crystallinity of the as-grown and annealed particles (Figure 3). The metathesis reaction scheme consistently produced amorphous Si NPs, as determined by their negligible XRD signal (Figure 3a) and a Raman stretch of 497 cm$^{-1}$ (Figure 3b), which is indicative of the broadened and shifted transverse-optical Raman mode associated with amorphous silicon (see Supporting

Information Table S1).[50] After annealing, strong XRD peaks were observed at positions agreeing with those corresponding to crystalline silicon standard, demonstrating that annealing provided a successful conversion from the amorphous to crystalline phase. The broadened and skewed Raman mode at 501 cm$^{-1}$ was also consistent with models for crystalline silicon NPs.[50] XRD peak analysis was used to determine the crystallite size. The average crystallite size from the Scherrer equation is 16.6 ± 3.0 nm, obtained from the three strongest diffraction peaks at 28.487 °, 47.395 ° and 56.194 °. The average crystallite size obtained by XRD measurements is larger than the average particle size obtained by TEM analysis due to the differing preparation methods of the samples for analysis. XRD provides an average crystallite size of the entire powder samples whereas the TEM grid is prepared from the sonicated solution. The larger NPs tend to precipitate rapidly and therefore the TEM grid presents a smaller average size than that observed by XRD.

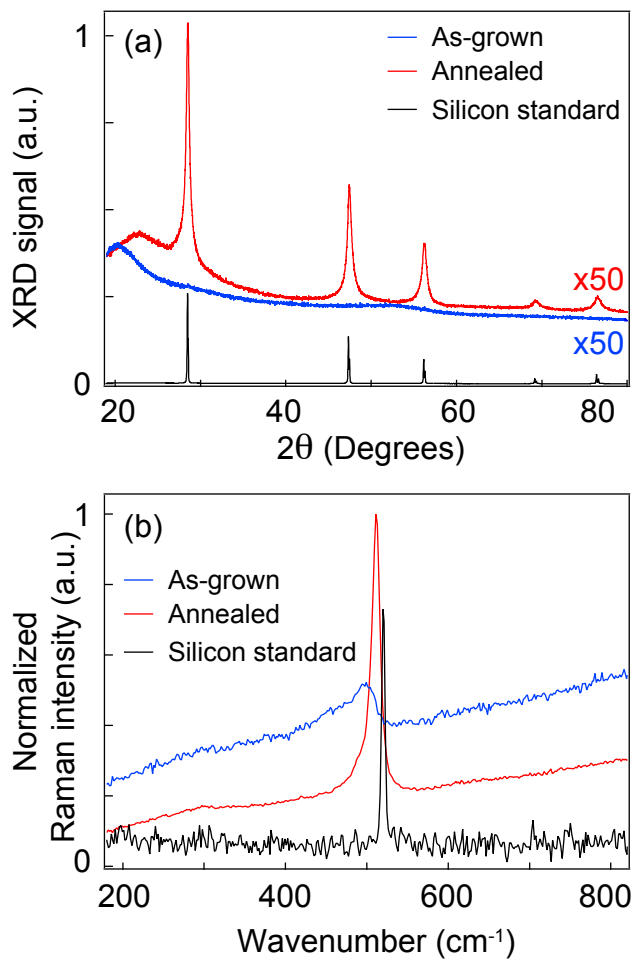

**Figure 3.** Crystallinity of as-grown and annealed Si NPs. (a) X-ray diffraction and (b) Raman spectroscopy of Si NPs before and after annealing show a transition from amorphous to crystalline states.

The surface passivation for each of the samples was determined by infrared spectroscopy (Figure 4). The octyl terminated Si NPs (Figure 4a) show stretches of the methylene groups at 2964 and 2891 cm$^{-1}$, together with a long chain methyl group observed at a wavelength of 796 cm$^{-1}$. The Si NPs passivated using 4-aminophenylmagnesium bromide show evidence of the amine group, indicated by a 3592 and 3517 cm$^{-1}$ stretch, together with the N-H bend and the N-H wag at 1639 cm$^{-1}$ and 802 cm$^{-1}$ respectively (Figure 4b). For Si NPs that were terminated with
9

APTMS, the weak secondary aliphatic amine stretch is observed at 3404 cm$^{-1}$ and the N-H bend and wag were observed at 1449 and 800 cm$^{-1}$ respectively. All samples showed evidence of oxidation, indicated by the Si-O-Si stretch at 1000-1100 cm$^{-1}$. This is typical for larger diameter Si NPs, as the decreased surface curvature reduces the surface coverage of the functional group. The frequency of the Si-O-Si stretch provides information of the structure of oxide layers as well as strain, porosity and oxygen content[51-53] and as the SiO$_2$ surface increases in thickness the stretching mode band moves towards higher frequencies. The higher frequency of the Si-O-Si stretch observed for the alkyl terminated Si NPs suggests a thicker oxide passivation for these particles compared to the APTMS terminated Si NPs.

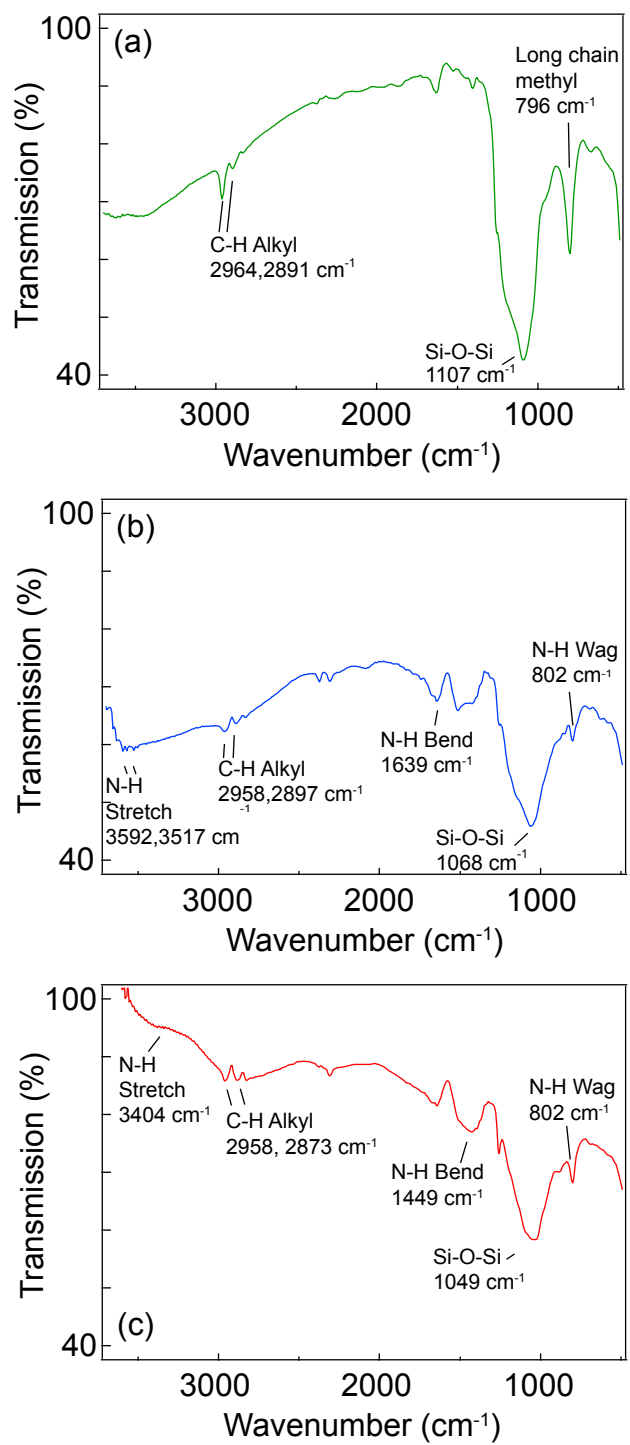

**Figure 4.** Infrared spectra of octyl, 4-aminophenyl, and APTMS terminated Si NPs. (a) Octyl terminated Si NPs, (b) 4-aminophenyl terminated Si NPs (c) APTMS terminated Si NPs.

Figure 5 shows the measured $^{29}$Si nuclear spin lattice relaxation times ($T_1$) of a range of synthesized Si NPs with different diameters. The $T_1$ and sizes of Si nano and microparticles measured in previous investigations are shown as a reference[32, 33] together with a spin diffusion based model for $T_1$ relaxation.[33] The particles synthesized in this study show long $T_1$ times, ~ 600 s for the annealed samples, similar to $T_1$ times measured for particles ~ 60 nm in diameter also synthesized by wet synthesis techniques.[32] For high purity Si particles the primary mechanism for relaxation is spin diffusion of the nuclear polarization, P, from the center of the particle to the particle surface, according to

$$\frac{dP}{dt} = -D\Delta P, \qquad (1)$$

where D is the spin diffusion constant, and so $T_1$ scales quadratically with particle diameter.[32, 53, 54] Unbonded electrons primarily at the silicon-silicon dioxide interface at the particle surface together with proximal spins contained within the passivating moiety can then drive local nuclear relaxation. The rate of this relaxation is set by the density of these spins and their own relaxation time, and scales with distance $r^{-6}$, where $r$ is the distance between the $^{29}$Si nucleus and relaxing spin.[36] Impurities within the particle such as free carriers, unbonded electrons or metal contaminants cause additional relaxation and shorten the $^{29}$Si nuclear $T_1$ time.[32, 53] Similar $T_1$ times were measured for the alkane and aromatic terminated particles, indicating that the choice of passivating moiety had little effect on the internal spin dynamics of the particle. We note that the $T_1$ times of the crystalline samples are shorter than the amorphous samples, potentially due to the introduction of isolated point defects in the silicon lattice brought about by the annealing process.

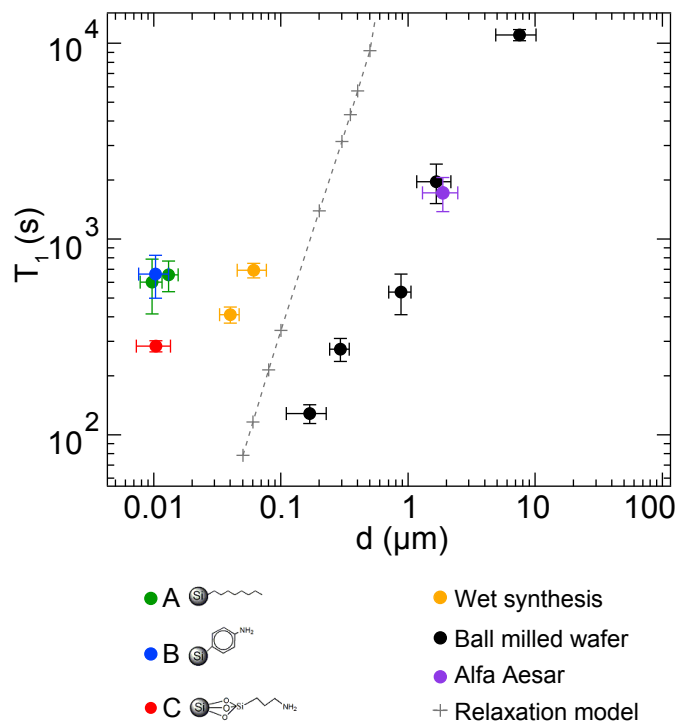

**Figure 5.** Nuclear spin relaxation ($T_1$) times at 2.9 T and 300 K as a function of particle diameter $d$ for Si nanoparticles synthesized by schemes described in Figure 1. Previously reported $T_1$ measurements of other silicon particles are shown as a reference. Simulations of nuclear relaxation based on a spin diffusion model are also shown. Vertical error bars are from exponential fits to relaxation data; horizontal error bars are from size distributions.

Natural oxidation of the surface of crystalline silicon creates a source of unbonded electrons that can be used for DNP.[55, 56] Electron spin resonance (ESR) was performed to determine the nature of the defects and their density. Figure 6a shows the weight-adjusted ESR spectra (B = 0.35 mT, T = 300 K) for each reaction scheme, while the frequency response of the $^{29}$Si nuclear polarization under DNP conditions (B = 2.89 T and T = 3.6 K) is shown in Figure 6b. For each sample, the ESR and DNP data are plotted across the same range of g-factors to aid in comparison. The g-factor extracted for each sample, as determined by the zero crossing in the ESR and DNP data, differs slightly (see Supporting Information Table S2), ranging from 2.0059 to 2.0087, however there is good agreement seen between the room temperature ESR and low temperature DNP data in each case. The defect density in these particles was lower than other particles previously investigated,[32, 33] ranging from $4e^{12} - 4e^{13}$ cm$^{-2}$ (see Supporting Information Table S2). The $P_b$ defects, which occur due to an oxygen vacancy at the interface between crystalline silicon and silicon oxide, have a g-factor that is strongly anisotropic[56] that is dependent on the orientation of the underlying crystalline lattice. For spherical nanoparticles, where one lattice orientation does not dominate over another, the orientation averaged g-factor for these defects is g = 2.006 which is in good agreement with the g-factor observed for our crystalline samples. Both amorphous samples show similar g-factors in the range 2.0078-2.0087. These are higher than the g = 2.005[57] and g = 2.0059[58] previously reported for bulk amorphous Si, potentially due to asymmetric strain.

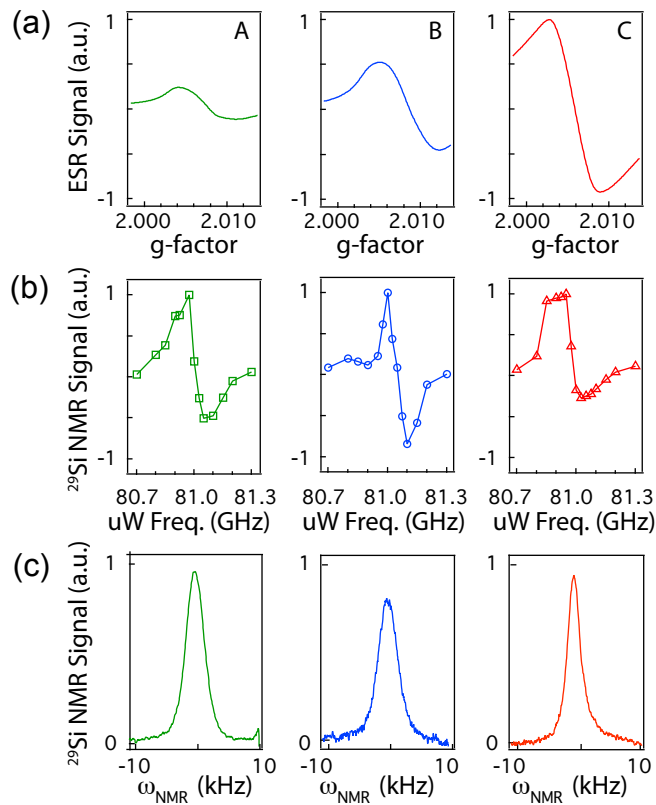

**Figure 6.** Electron spin resonance and dynamic nuclear polarization of A-octyl, B-4-aminophenyl and C-APTMS terminated Si NPs as indicated in the legend for Figure 5. (a) Electron spin resonance signal and (b) $^{29}$Si nuclear polarization as a function of microwave frequency under DNP conditions. (c) $^{29}$Si NMR spectra obtained under DNP conditions.

All samples show a positive enhancement in nuclear polarization at microwave frequencies below the ESR frequency and negative polarization enhancements at frequencies above it, indicating that the polarization is generated by a dipolar interaction rather than a contact hyperfine interaction between the unbonded electrons and nearby $^{29}$Si nuclei. The width of the DNP response is significantly narrower for the amorphous samples when compared to the crystalline samples, indicating that the polarization mechanism may be due to a single electron single nucleus dipole-dipole interaction (commonly referred to as the Solid Effect)[36] for the amorphous samples, rather than a multi-electron single nucleus dipole-dipole interaction which occurs in the crystalline samples (commonly referred to as Thermal Mixing).[55] The $^{29}$Si NMR spectra for the three samples after 3 h of DNP polarization are shown in Figure 6c. The amorphous Si NP samples show a broad resonance with a linewidth (full width at half maximum) of ~ 3.6 kHz. The crystalline Si NP sample shows a slightly narrower resonance (linewidth ~ 2.2 kHz) shifted by ~ 600 Hz from the amorphous peak. Both linewidths are broader than the ~ 500 Hz linewidth observed for bulk crystalline silicon[55] and 1.8 kHz linewidth observed for bulk amorphous silicon,[58] most probably due to strain.

The time evolution of the $^{29}$Si nuclear polarization under DNP conditions for each of the samples is shown in Figure 7a. For each measurement, the microwave frequency was set to the peak value in Figure 6b before a saturation recovery sequence performed and the $^{29}$Si NMR signal measured with a single 90° pulse. All polarization curves are well fit by a biexponential function,

$$P = P_\infty + (P_0 - P_\infty)\left(\alpha e^{\left(-t_{pol}/T_{1f}\right)} + (1-\alpha)e^{\left(-t_{pol}/T_{1s}\right)}\right), \quad (2)$$

with a short time constant $T_{1f}$ ~ 400 s, and a longer time constant $T_{1s}$ ranging from 3200 – 5700s (see Supporting Information Table S3). Here $\alpha$ corresponds to the fraction of nuclei that are being polarized directly *via* the dipole-dipole interaction and *(1-$\alpha$)* the fraction that are polarized indirectly *via* nuclear spin diffusion. Both the amorphous samples show $\alpha=0.13$, where for the annealed crystalline sample $\alpha=0.23$, which is due to the higher concentration of unbonded electrons in this sample (see Supporting Information Table S2). Figure 7b shows the decay in $^{29}$Si nuclear hyperpolarization measured at 3.6 K for each of the Si NP samples. The samples were polarized for 6 h with microwave irradiation at the peak value in Figure 6b, then the microwaves were then turned off and the hyperpolarization allowed to decay. The data are well fit by a single exponential,

$$P = P_0 + (P_f - P_0)e^{\left(-t_{dep}/T_{1,dep}\right)},$$ (3)

with all samples showing a depolarization time constant $T_{1,dep}$ greater than 300 s (see Supporting Information Table S3). The octyl terminated Si NPs showed the longest depolarization time constant of ~ 600 s, consistent with the lower density of electronic defects observed. The long depolarization times of the amorphous Si NPs are surprising, as previous DNP investigations of Si microparticles composed of amorphous and crystalline silicon showed relaxation of the amorphous component within seconds after microwave irradiation was ceased.[55]

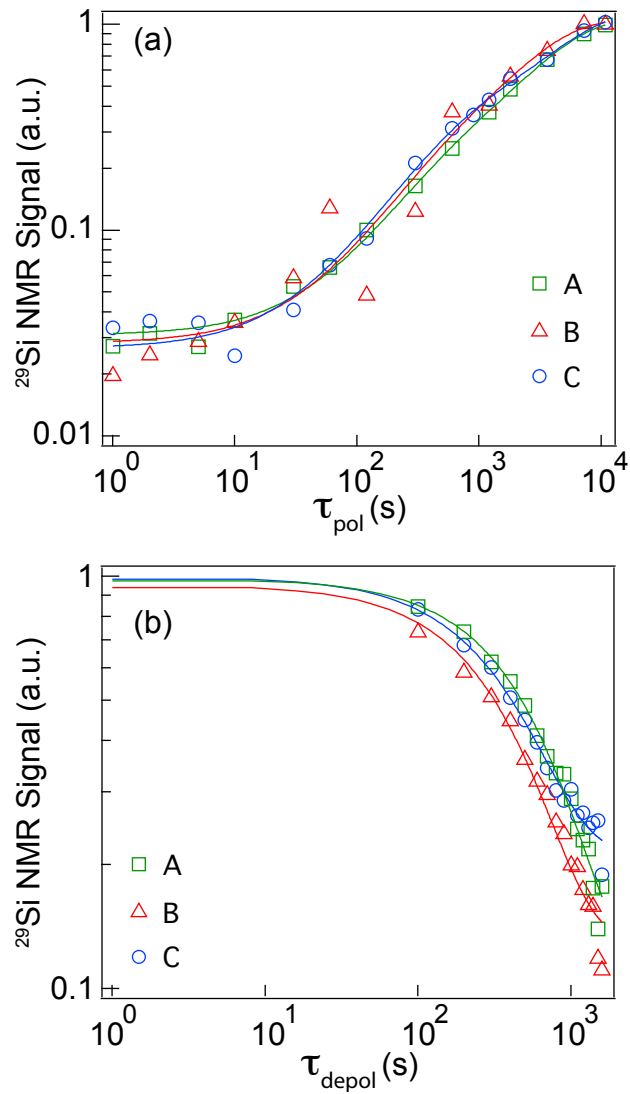

**Figure 7.** Time dependence of $^{29}$Si polarization measured at 3.6 K in (A) octyl, (B) 4-aminophenyl and (C) APTMS terminated Si NPs synthesized according the scheme described in Figure 1. (a) $^{29}$Si NMR signal as a function of polarization time $\tau_{pol}$ (b) Depolarization of the $^{29}$Si nuclear polarization at 3.6 K after 6 hours of DNP. The solid lines show fits to the data described in the text.

CONCLUSION

We have successfully produced Si NPs having long $^{29}$Si nuclear $T_1$ times with an average size of ~ 10 nm using a metathesis reaction of sodium silicide with a silicon tetrachloride. Combining these Si NPs with Grignard reagents resulted particles that were partially passivated with alkyl and aromatic ligands. The partial surface passivation allows sufficient oxidation to provide hyperpolarization sites for DNP without compromising the non-aggregating characteristics of the sample. Additionally, the as-grown amorphous Si NP samples could be annealed into crystalline Si NPs that retained their nanoscale size and did not lead to larger aggregates due to sintering. Both the amorphous and crystalline Si NPs showed $T_1$ times more than two orders of magnitude larger than diamond nanoparticles of similar sizes, and a 5 fold improvement on size-adjusted $T_1$ times when compared to silicon particles previously studied.[32] All Si NPs synthesized in this study could be hyperpolarized by low temperature dynamic nuclear polarization, and retained their polarization for > 300 s, demonstrating that the long $T_1$ times of Si NPs persist down to the nanoscale.

EXPERIMENTAL METHODS

**Chemicals.** Silicon tetrachloride ($SiCl_4$, 99.998 %) was purchased from Aldrich and was distilled. Silicon tetraiodide ($SiI_4$, 99 %) and silicon tetrabromide ($SiBr_4$, 99%) were purchased from Alfa Aesar and were used without further treatment. Sodium hydride powder (NaH, 95 %) and silicon powder (Si, 99 %) were purchased from Sigma-Aldrich and were used without further treatment. The aromatic halide 4-bromoaniline ($C_6H_6BrN$, 97 %) and the alkyl halide octyl bromide ($C_8H_{17}Br$, 99 %) were purchased from Aldrich and were used without further treatment. Tetraethylorthosilicate (TEOS, 99.999%) and poly(ethylene glycol) methyl ether (PEG) were purchased from Aldrich. Aminopropyltrimethoxysilane (APTMS, 97%) and dioctylether (DOE, 99%), dimethoxyethane (DME, 99%) and squalane ($C_{30}H_{62}$) were purchased from Aldrich. All manipulations were carried out under dry $N_2$, in either a glove box or a Schlenk line, using standard anaerobic and anhydrous techniques.

**Preparation of Na$_4$Si$_4$ precursor.** Sodium silicide, Na$_4$Si$_4$, was prepared according to a modified previously published procedure.[59] A Spex 8000 M ball mill with a Spex tungsten carbide vial and two tungsten carbide balls (diameter of ~1 cm) were used to ball-mill mixtures of NaH and Si. The intimately mixed mixture of NaH and Si was placed into a 1 mL alumina crucible with a 1.5 mL alumina boat as a cover. The crucibles were placed into a quartz tube with rubber stopcocks on both ends. The quartz tube was removed from the glove box and placed in a horizontal tube furnace and connected to Ar flowing at 30 mL/min. The quartz tube was heated at 420°C for 48h. After cooling to room temperature a black solid was isolated and stored in a glove box.

*CAUTION*: NaH and Na$_4$Si$_4$ are highly reactive to air and moisture and must be handled under an inert atmosphere.

**Synthesis of halide terminated Silicon NPs *via* Sodium Silicide with Silicon tetrachloride.** The Na$_4$Si$_4$ (1 mmol) and SiCl$_4$ (2 mmol) were placed into a 3-neck 1000 mL round bottom flask in a dry box. The solvent DME, DOE or DOE/squalane was deoxygenated and 400 mL of DME, DOE or DOE/squalane was added to the starting reagents by a cannula. The solution was allowed to reflux overnight and the solution was allowed to cool. The solvent was removed *via* short bridge under vacuum and 200 mL of anhydrous THF was added to the Schlenk flask and stirred.

**Synthesis of 4-aminophenyl terminated Silicon NPs.** To obtain magnesium, 4-aminophenyl bromo (C$_6$H$_6$NMgBr) a Grignard reaction was performed by reacting 1.3 moles of 4-Bromoaniline with 2 moles of magnesium and refluxed overnight. The Si NPs in THF was stirred and 2.0 mL of 0.5M 4-aminophenylmagnesium bromide (C$_6$H$_6$NMgBr) in tetrahydrofuran (THF) was added dropwise. The Grignard reagent was added over a ten minute period. The mixture was heated and allowed to reflux overnight. The reaction was allowed to cool and the solution was separated by centrifugation (8000 rpm for 20 minutes). The liquid layer was decanted and the solid was placed into a separatory funnel with a water hexane layer and the suspended layer of solid between the water/hexane layer was isolated. The solid was washed three times with a total of 60 mL of methanol and centrifuged. The yield of product obtained was approximately 30%.

**Synthesis of octyl terminated Silicon NPs.** To obtain octylmagnesium bromide a Grignard reaction was performed by reacting 1.3 moles of octylbromide with 2 moles of magnesium and refluxed overnight. The Si NPs in THF was stirred and 2.2 mL of 0.5M octylmagnesium bromide ($C_8H_{17}MgBr$) in THF was added dropwise. The Grignard reagent was added over a ten minute period. The mixture was heated and allowed to reflux overnight. The reaction was allowed to cool and the solution was separated by centrifugation (8000 rpm for 20 minutes). The liquid layer was decanted and the solid was placed into a separatory funnel with a water hexane layer and the suspended layer of solid between the water/hexane layer was isolated. The solid was washed three times with a total of 60 mL of methanol and centrifuged. A yield of approximately 30% was obtained.

**Synthesis of Silicon NPs *via* Sodium Silicide with silicon tetrachloride, annealing and aminopropyltrimethoxysilane (APTMS) coating.** The coating was a modified procedure previously published in literature.[60] The washed solid obtained from the synthesis of sodium silicide with $SiCl_4$ was placed into 5 mL of methanol and allowed to stir overnight. The solvent was evaporated by rotary evaporation and was annealed at 650°C for 20 minutes under an 80:20 $H_2/N_2$ gas flow mixture in a flow furnace with a flow rate of 30 mL/min. The solid was placed back into fresh DME and 5 mL of H$_2$O was added and allowed to stir overnight followed by the addition of 0.3 mL (1.8 mmol) of aminopropyltrimethoxysilane and the resulting solution was stirred overnight. The solvent was evaporated under reduced pressure. The solid was washed with HPLC H$_2$O to remove any un-reacted APTMS.

**Characterization.** Powder X-ray diffraction (XRD) data was collected using an air tight holder on a Bruker D8 diffractometer operating at 40 kV and 40 mA with CuKα radiation (λ = 1.54184 Å). Morphology and chemical functionality were analyzed by a CM-12 transmission electron microscope (TEM) with an accelerating voltage of 120 kV. Samples were ground with a 1:5 ratio of Si NP sample and anhydrous KBr and diffuse reflectance was performed using a Shimadzu IR Prestige-21 Fourier transform infrared spectrophotometer (FTIR). High resolution TEM (HRTEM) was performed on a JEOL 2500SE Schottky emitter microscope operating at 200kV and equipped with a Gatan multiscan camera. TEM samples were prepared by dissolving the silicon NPs into isopropanol and dipping the grids into the solution. For HRTEM, a drop of the solution was put on the copper grid and dried overnight under a heat lamp. Raman spectroscopy was performed on a Renishaw RM1000 laser Raman microscope equipped with Argon 514 nm laser for excitation guided by a microprobe.

**Electron Spin Resonance.** Continuous wave x-band electron spin resonance measurements were performed on the samples at room temperature (Bruker ElexSys E500). The a.c. field (amplitude 0.001 mT, f$_{mod}$ = 100 kHz) was swept from 343.5 mT to 358.5 mT over a period of 60 s. An estimate of the electron spin concentration on the particle surface was made by taking ESR spectra with a piece of phosphorus-doped silicon wafer also inserted in the spectrometer cavity. The addition of the wafer piece resulted in the appearance of additional ESR signal due to the n-type doping. Comparing the peak areas and using the known doping level of the wafer piece (0.008 − 0.01 Ω.cm) yielded an order-of-magnitude estimate of the mean volume density of defect spins.

**Room Temperature $^{29}$Si $T_1$.** The room temperature $T_1$ of the samples were measured at B = 2.89 T with a saturation recovery Carr-Purcell-Meiboom-Gill (CPMG) sequence. A train of 32 hard $\pi/2$ pulses saturated residual magnetization, and the sample was allowed to polarize for a time τ. The magnetization was rotated to the transverse plane with a single $\pi/2$ pulse and (N=200) π pulses performed. The echos following each of the π pulses were Fourier transformed and the height of the peak was used as a measure of the nuclear polarization. The data were fitted with an exponential function $P = P_0 + Ae^{-\tau/T_1}$ and the characteristic polarization time constant ($T_1$) extracted.

**$^{29}$Si Dynamic Nuclear Polarization.** Low temperature microwave induced dynamic nuclear polarization (DNP) was performed at T = 3.6 K using a home-built DNP polarizer operating at B = 2.89 T, $f_{ESR}$ = 81 GHz. Details of the polarizer construction are described in Ref. 33. A saddle coil that could be tuned across a wide range of temperatures was used for NMR detection. The $^{29}$Si nuclear polarization was measured in-situ under DNP using a saturation recovery sequence. A train of 32 hard $\pi/2$ pulses saturated residual magnetization, and the sample was allowed to polarize for a time τ. The magnetization was rotated to the transverse plane with a single $\pi/2$ pulse and the free induction decay measured. The area under the Fourier transformed peak was used as a measure of the $^{29}$Si nuclear polarization. For depolarization measurements, the samples were polarized at the frequency giving the highest nuclear polarization for 6 h. The microwaves were then turned off and the remaining $^{29}$Si polarization measured with series of 20 degree pulses separated by 100 s. A mathematical correction was

applied in post processing to take into account the excess magnetization used up by earlier pulses in the sequence.

**Supporting Information**. Image of suspended NPs, dynamic nuclear polarization and electron paramagnetic resonance data (pdf, 2 pages). This material is available free of charge *via* the Internet at http://pubs.acs.org.

AUTHOR INFORMATION

**Corresponding Authors**

*smkauzlarich@ucdavis.edu (SMK) and *marcus@nbi.dk (CMM).

ACKNOWLEDGMENT

We recognize financial support from DOE (DESC0002289), NIH (1 R21 EB007486-01A1), the NSF BISH Program (CBET-0933015), and the Harvard NSF Nanoscale Science and Engineering Center.

TOC/Abstract Graphic

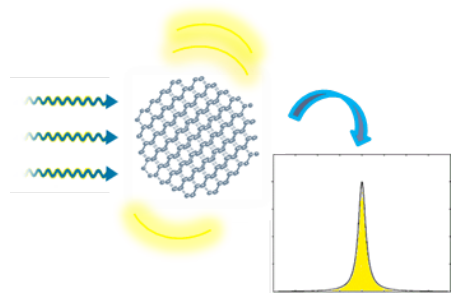